\documentclass[prl,
twocolumn,
notitlepage,
preprintnumbers,
nofootinbib,APS, %10pt,
superscriptaddress]{revtex4}

\usepackage{graphics,epsfig,epsf,amsmath,amssymb,amsfonts,latexsym,cancel}
\usepackage[english]{babel}
\usepackage{float}
\usepackage{multirow}
\usepackage{array}
\usepackage{dcolumn}
\usepackage[usenames]{color}
\usepackage{color,bm}
\usepackage[colorlinks=true, linkcolor=navyblue, urlcolor=navyblue, citecolor=darkred]{hyperref}
\definecolor{navyblue}{rgb}{0, 0.0, 1.0}
\definecolor{darkred}{rgb}{1.0, 0.0, 0.0}
\newcommand{\vecc}[1]{\mbox{\boldmath $#1$}}
\def\ba{\begin{eqnarray}}
\def\ea{\end{eqnarray}}
\def\be{\begin{equation}}
\def\ee{\end{equation}}
\def\GEV{\mbox{GeV}}
\def\bc{\begin{center}} % ---------------
\def\ec{\end{center}}   % -------------------
\def\nn{\nonumber}

\begin{document}

\title{
New Method for Measuring the Ratio $\mu_p G_E/G_M$ Based on the Polarization Transfer
from the Initial Proton to the Final Electron in the $e \vec p \to \vec e  p$ Process}

\author{M.\,V.~Galynskii}
\email{galynski@sosny.bas-net.by}
\affiliation{Joint Institute for Power and Nuclear Research -- Sosny,
        National Academy of Sciences of Belarus, 220109 Minsk, Belarus}

\author{Yu.\,M.~Bystritskiy}
\email{bystr@theor.jinr.ru}
\affiliation{Joint Institute for Nuclear Research, 141980 Dubna, Moscow Region, Russia}

\author{V.\,M.~Galynsky}
\affiliation{Belarusian State University, Minsk 220030, Belarus}
%\email{galynsky@bsu.by}

\begin{abstract}
In this letter, we %have proposed
propose a new method for measuring the Sachs form factors ratio
($R =\mu_p G_E/G_M$) based on the transfer of polarization from the initial proton to the final
electron in the elastic $e \vec p \to \vec e p$ process, in the case when the axes of quantization
of spins of the target proton at rest and of the scattered electron are parallel, i.e., when an electron
is scattered in the direction of the spin quantization axis of the proton target.
To do this, in the kinematics of the SANE collaboration experiment (2020) on measuring double spin
asymmetry in the $\vec e\vec p \to e p$ process, using Kelly (2004) and Qattan (2015) parametrizations,
a numerical analysis was carried out of the dependence of the longitudinal polarization degree
of the scattered electron on the square of the momentum transferred to the proton, as well
as on the scattering angles of the electron and proton.
It is established that the difference in the longitudinal polarization degree of the final electron
in the case of conservation and violation of scaling of the Sachs form factors can reach 70\%.
This fact can be used to set up polarization experiments of a new type to measure the ratio $R$.
\end{abstract}

\maketitle

{\it Introduction}.---Experiments on the study of electric ($G_{E}$) and magnetic ($G_{M}$) proton
form factors, the so-called Sachs form factors (SFF), have been conducted since the mid-1950s
\cite{Hofstadter1958} in the process of elastic scattering of unpolarized electrons off a proton.
At the same time, all experimental data on the behavior of SFF were obtained using the Rosenbluth
technique (RT) based on the use of the Rosenbluth cross section (in the approximation of the one-photon
exchange) for the $ep \to ep$ process in the rest frame of the initial proton \cite{Rosen}
\ba
\label{Ros}
\frac{d\sigma} {d\Omega_e}=
\frac{\alpha^2E_2\cos^2(\theta_e/2)}{4E_1^{3}\sin^4(\theta_e/2)}
\frac{1}{1+\tau_p} \left(G_E^{2}
+\frac{\tau_p}{\varepsilon}\,G_M^{2}\right).
\ea
Here $\tau_p=Q^2/4m^2$, $Q^2=4E_1 E_2\sin^2(\theta_e/2)$ is the square
of the 4-momentum transferred to the proton; $m$ is the mass of the proton; $E_1$ and $E_2$
are the energies of the initial and final electrons; $\theta_e$ is the electron scattering angle;
$\varepsilon=[1+2(1+\tau_p)\tan^2(\theta_e/2)]^{-1}$ is the degree of linear (transverse)
polarization of the virtual photon \cite{Dombey,Rekalo74,AR,GL97};
and $\alpha=1/137$ is the fine structure constant.

For large values of $Q^2$, as follows from formula (\ref{Ros}), the main contribution to the cross
section of the $ep\to ep$ process is given by the term proportional to $G_M^{\,2}$, which is already
at $Q^2\geqslant 2~\GEV^2$ leads to significant difficulties in extracting the contribution of
$G_E^{\,2}$ \cite{ETG15,Punjabi2015}.

With the help of RT, the dipole dependence of the SFF on $Q^2$ in the region $Q^2 \leqslant 6~\GEV^2$ was established
\cite{ETG15,Punjabi2015}. As it turned out, $G_E$ and $G_M$ are related by the
scaling ratio $G_M\approx\mu_p G_E$ ($\mu_p=2.79$ -- the magnetic moment of the proton),
and for their ratio $R \equiv\mu_p G_E/G_M$, the approximate equality $R \approx 1$ is valid.

In the paper of Akhiezer and Rekalo \cite{Rekalo74}, a method for measuring the ratio of $R$ is proposed
based on the phenomenon of polarization transfer from the initial electron to the final proton in
the $\vec e p\to e\vec p$ process. Precision JLab experiments \cite{Jones00,Gay01,Gay02}, using this method,
%conducted in the Jefferson Laboratory (JLab, USA)
found a fairly rapid decrease in the ratio of $R$ with an increase in $Q^2$, which indicates
a violation of the dipole dependence (scaling) of the SFF. In the range
$0.4~\GEV^2 \leqslant Q^2 \leqslant 5.6~\GEV^2$, as it turned out, this decrease is linear.
Next, more accurate measurements of the ratio $R$ carried out in
\cite{Pun05,Puckett10,Puckett12,Puckett17,Qattan2005} in a wide area in $Q^2$ up
to $8.5~\GEV^2$ using both the Akhiezer--Rekalo method \cite{Rekalo74} and
the RT \cite{Qattan2005}, only confirmed the discrepancy of the results.

In the SANE collaboration experiment \cite{Liyanage2020} (2020), the values of $R$ were obtained
by the third method \cite{Dombey,Donnelly1986} by extracting them from the results
of measurements of double spin asymmetry in the $\vec e\vec p\to e p$ process in the case,
when the electron beam and the proton target are partially polarized.
The extracted values of $R$ in \cite{Liyanage2020} are consistent with the experimental results
\cite{Jones00,Gay01,Gay02,Pun05,Puckett10,Puckett12,Puckett17}.

In \cite{JETPL2008,JETPL18,JETPL19,JETPL2021,PEPAN2022,JETPL2022}, the 4th method
of measuring $R$ is proposed based on the transfer of polarization from the initial
proton to the final one in the $e\vec p\to e\vec p$ process in the case when their spins
are parallel, i.e. when the proton is scattered in the direction of the quantization axis
of the spin of the resting proton target.

In this paper, the 5th method of measuring the ratio of $R$ is proposed based on the transfer
of polarization from the initial proton to the final electron in the process $e\vec p\to\vec e p$
in the case when their spins are parallel, i.e. when the electron is scattered in the direction of
the spin quantization axes of the resting proton target.

{\it The helicity and diagonal spin bases}.---The spin 4-vector $s=(s_{0}, \vecc s)$
of the fermion with 4-momentum $p$ ($p^2=m^2)$ satisfying the conditions of orthogonality
($sp=0$) and normalization ($s^{2} = - 1$), is given by
\ba
s=(s_{0}, \vecc s), \quad
s_{0}=\frac {\vecc c \,\vecc p}{m}, \quad
\vecc s =\vecc c + \frac{\,(\vecc c \,\vecc p)\,\vecc p} {m(p_0 +m )\,}\;,
\label{spinq}
\ea
where %the 3-vector
$\vecc c$ ($\vecc c^{2}=1$) is the axis of spin quantization.

Expressions (\ref{spinq}) allow us to determine the spin 4-vector $s=(s_{0}, \vecc s)$
by a given 4-momentum $p=(p_{0}, \vecc p)$ and 3-vector $\vecc c$.
On the contrary, if the 4-vector $s$ is known, then the spin quantization axis $\vecc c$
is given by %calculated by the formula
\ba
\vecc c = \vecc s - \frac{s_0}{p_0+m} \, \vecc p,
\label{spinc}
\ea
i.e. %the vectors
$\vecc c$ and $s$ for a given $p$ uniquely define each other.

At present, the most popular in high-energy physics is the helicity basis \cite{Jacob}, in which
the spin quantization axis is directed along the momentum of the particle
($\vecc c =\vecc n = \vecc{p}/|\vecc p|$), while the spin 4-vector (\ref{spinq}) %has the form
defined as
\ba
s=(s_{0}, \vecc s) =(|\vecc v |, v_{0} \, \vecc n),
\label{spins}
\ea
where $v_{0}$ and $\vecc v$ are the time and space components
of the 4-velocity vector $v=p/m$ ($v^2=1$).

For the process under consideration
\ba
e(p_1)+ p\,(q_1,s_{p_{1}}) \to e(p_2,s_{e_{2}})+ p \,(q_2),
\label{EPEPpe}
\ea
where $p_1$, $q_1$ ($p_2$, $q_2$) are the 4-momenta of the initial (final) electrons
and protons with masses $m_0$ and $m$, it is possible
to project the spins of the initial proton and the final electron in one
common direction %determined
given by %the 3-vector
\cite{FIF70,GL}
\ba
\label{osq1p2}
\vecc a = \vecc p_{2}/p_{20} - \vecc q_{1}/q_{10}.
\ea
Since the common axis of spin quantization (\ref{osq1p2}) defines the spin basis
and is the difference of two three-dimensional vectors, the geometric image of which
is the diagonal of the parallelogram, it is natural to call
it the diagonal spin basis (DSB).
In it, the spin 4-vectors of the initial proton $s_{p_{1}}$
and the final electron $s_{e_{2}}$ are given by % have the form:
\ba
\label{DSBpr1}
s_{p_{1}}&=&\frac { m^2 p_{2} - ( q_{1} p_{2})q_{1} }
{m\sqrt{ ( q_{1}p_{2} )^{2} - m^2 m_0^2 }},\\
\label{DSBel2}
s_{e_{2}}&=& \frac { ( q_{1} p_{2}) p_{2} - m_0^2 q_{1} }
{m_0\sqrt{ ( q_{1}p_{2} )^{2} - m^2 m_0^2 }}. %\nn
\ea
Note that in the papers \cite{JETPL2008,JETPL18,JETPL19,JETPL2021,PEPAN2022,JETPL2022}
was used the analogous DSB for the initial and final protons (with a common spin quantization axis
$\vecc a = \vecc q_{2}/q_{20} - \vecc q_{1}/q_{10}$) \cite{Sik84,GS98}.

In the laboratory %reference
frame (LF), where the initial proton rests,
$q_1=(m,\vecc 0)$, the spin 4-vectors (\ref{DSBpr1}), (\ref{DSBel2}) %take the form:
reduces to
\ba
\label{DSB_LSO_q1p2}
s_{p_{1}}=(0, \vecc {n_{2}} )\,, \; s_{e_{2}}= (|\vecc {v_{2}}|, v_{20}\, \vecc {n_{2}} )\,,
\ea
where $\vecc n_{2}= \vecc {p_2}/|\vecc p_2|$, $v_{2}=(v_{20}, \vecc {v_{2}})=p_2/m_0$.

Using the explicit form of the spin 4-vectors $s_{p_{1}}$ and $s_{e_{2}}$ (\ref{DSB_LSO_q1p2})
and formulas (\ref{spinc}) or (\ref{osq1p2}), it is easy to verify that the quantization axes
of the initial proton and the final electron spins in the LF have the same form and coincide
with the direction of the final electron momentum
\ba
\vecc a=\vecc c = \vecc c_{p_{1}} =\vecc c_{e_{2}}=\vecc n_{2}= \vecc {p_2}/|\vecc p_2|\,.
\label{osq1p2222}
\ea

In the ultrarelativistic limit, when the electron mass can be neglected
(i.e. at $p_{10}, p_{20}\gg m_0$), the spin 4-vectors (\ref{DSBpr1}), (\ref{DSBel2}) reduces to
\ba
s_{p_{1}} = \frac { m^2 p_{2} - ( q_{1} p_{2})\,q_{1} }
{m ( q_{1}p_{2} )}, \qquad s_{e_{2}} = \frac {p_2}{ m_0} \, .
\label{DSBq1p2m0}
\ea

Below, in the ultrarelativistic limit, we present the main kinematic relations
used in conducting numerical calculations of polarization effects in
the $e\vec p\to\vec e p$ process in the LF.

{\it Kinematics}.---The energies of the final electron $E_2$ and proton $E_{2p}$ are connected in the LF with
the square of the momentum transferred to the proton $Q^2=-q^2$, $q^2=(q_2-q_1)^2$ as follows
\ba
\label{E2Q}
&&E_2=E_1-Q^2/2m, \qquad E_{2p}=m+Q^2/2m, \\
\label{E2tau}
&&E_2=E_1-2m\tau_p, \qquad E_{2p}=m(1+2\tau_p).
\ea

The dependence of $E_2$ and $Q^2$ on the scattering angle of the electron $\theta_e$ in the LF has the form
\ba
\label{E2tea}
E_2(\theta_e)&=&\frac{E_1}{1+ (2E_1/m)\, \sin^2 (\theta_e/2)}, \\
\label{Q^2te}
Q^2(\theta_e)&=&\frac {4 E_1^{\,2} \sin^2 (\theta_e/2)}{1+ (2E_1/m) \sin^2 (\theta_e/2)},
\ea
where $\cos(\theta_e)=\vecc p_1 \vecc p_2/|\vecc p_1| |\vecc p_2|$.

The dependence of $E_{2p}$ and $Q^2$ on the scattering angle of the proton $\theta_p$
in the LF has the form
\ba
\label{E2tp}
E_{2p}(\theta_p)&=&m\, \frac{(E_1+m)^2+E_1^{\,2}\cos^2(\theta_p)}
{(E_1+m)^2-E_1^{\,2}\cos^2(\theta_p)}\,,\\
\label{Q^2tp}
Q^2(\theta_p)&=&\frac{4m^2 E_1^{\,2}\cos^2(\theta_p)}
{(E_1+m)^2-E_1^{\,2}\cos^2(\theta_p)}\,,
\ea
where $\cos(\theta_p)=\vecc p_1 \vecc q_2/|\vecc p_1| |\vecc q_2|$.

The dependence of the scattering angles $\theta_e$ and $\theta_p$ on $E_1$ and $Q^2$ has the form
\ba
&&\label{tetae}
\theta_e=\arccos{\left(1-\frac{Q^2}{2 E_1 E_2} \right)}, \\
\label{tetap}
&& \theta_p=\arccos{\left(\frac{E_1+m}{E_1} \sqrt{\frac{\tau_p}{1+\tau_p}}\;\right)}.
\ea

In the elastic $ep\to ep$ process an electron can be scattered by an angle of
$0^{\circ} \leqslant \theta_e\leqslant 180^{\circ}$, while the scattering angle of the proton
$\theta_p$ varies from $90^{\circ}$ to $0^{\circ}$ \cite{AR}. Possible values of $Q^2$ lie in
the range $0 \leqslant Q^2\, \leqslant Q^2_{max}$, where
\ba
\label{Q2max}
Q^2_{max}=\frac{4mE_1^{\,2}}{(m+2E_1)} \,.
\ea

\begin{figure}%[h!]
\centering
\includegraphics[width=0.45\textwidth]{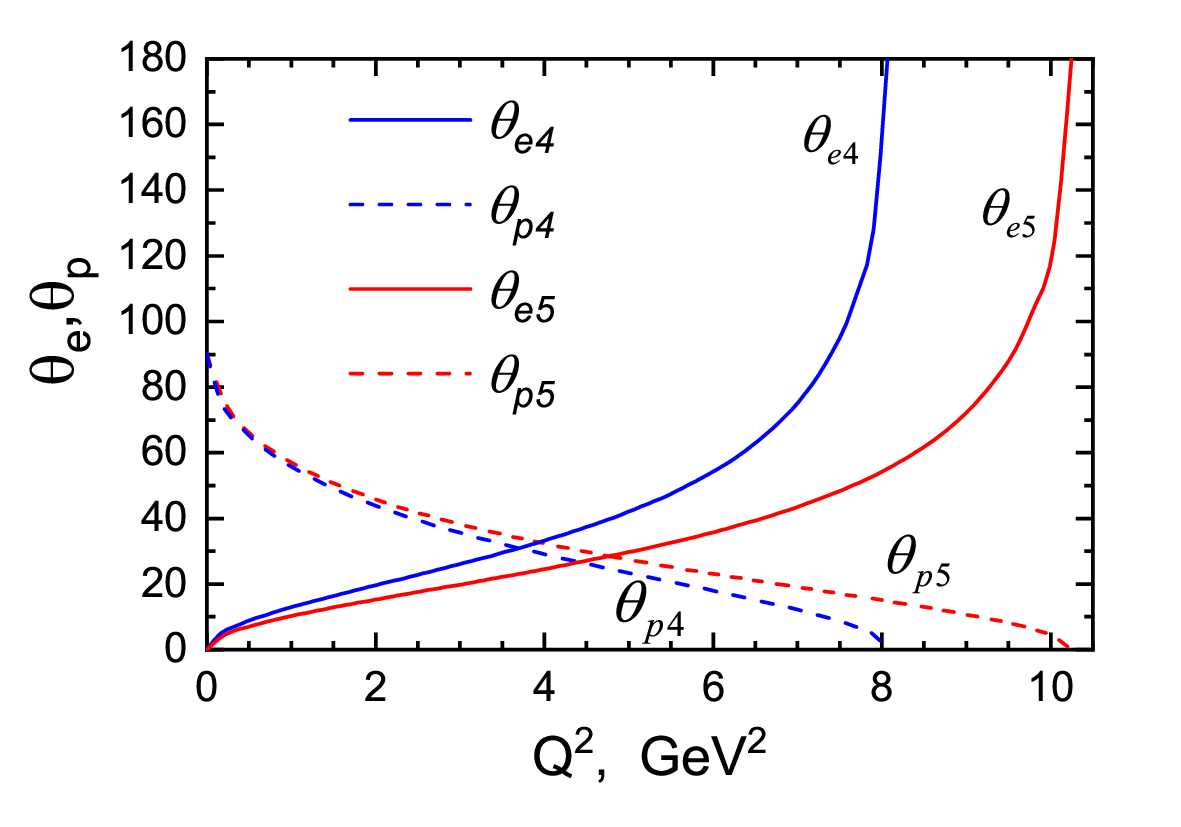}
\vspace{-4mm}
\caption{
$Q^2$-dependence of the scattering angles of the electron $\theta_e$ and the proton $\theta_p$
(in degrees) at electron beam energies in the %SANE collaboration
experiment \cite{Liyanage2020}. The lines $\theta_{e4}$, $\theta_{p4}$ ($\theta_{e5}$, $\theta_{p5}$)
correspond to $E_1=4.725$ ($5.895$)~$\GEV$.
}
\label{Theta_ep1}
\end{figure}

The results of calculations of the dependence of the scattering angles of the electron and proton
on the square of the momentum transferred to the proton $Q^2$ in the $e p \to e p$ process
at electron beam energies $E_1=4.725$ and $5.895~\GEV$ in the SANE collaboration experiment
\cite{Liyanage2020} are presented by the graphs in Figure~\ref{Theta_ep1}. They correspond
to lines with labels $\theta_{e4}$, $\theta_{p4}$ and $\theta_{e5}$, $\theta_{p5}$.
%\vspace{-1mm}

%\vspace{-4mm}
Information about the electron and proton scattering angles (in radians) at electron
beam energies and values of $Q^2$ in the %SANE collaboration
experiment \cite{Liyanage2020}
is presented in Table~\ref{Uglyep}. It also contains the corresponding values
for $Q^2_{max}$ (\ref{Q2max}).

\vspace{-5mm}
\begin{table}[h!]
\centering
\caption{
The scattering angles of the electron $\theta_{e}$ and proton $\theta_{p}$ (in radians) at
electron beam energies $E_1=4.725$ and $5.895~\GEV$ and values $Q^2$ equal to $2.06$ and $5.66~\GEV^2$.
\vspace{0.8mm}
}
\label{Uglyep}
\tabcolsep=1.70mm
\footnotesize
\begin{tabular}
{| c | c | c | c | c | c | }
\hline
$E_1$ ($\GEV$)
&  $Q^2$ ($\GEV^2$)
& $\theta_{e}(rad)$ & $\theta_{p}(rad)$
& $Q^2_{max}$ ($\GEV^2$)
  \\
\hline
5.895 &2.06 & 0.27 &  0.79  & 10.247  \\
\hline
5.895 &5.66 & 0.59 & 0.43  & 10.247 \\
\hline
4.725 &2.06 & 0.35 & 0.76  & 8.066  \\
\hline
4.725 &5.66 & 0.86 & 0.35 & 8.066 \\
\hline
\end{tabular}
\end{table}

{\it Cross section of the $e\vec p \to \vec e p$ process}.---In the one-photon exchange
approximation, the differential cross section of the process (\ref{EPEPpe}),
calculated in an arbitrary reference frame in the DSB (\ref{DSBpr1}), (\ref{DSBel2}), reads
\ba
\label{T2all}
\frac {d \sigma_{e \vec p \to \vec e p }}{d t}&=&\frac{ \pi \alpha^2 }
{2\lambda_s (1+\tau_p)}\,\frac {|{T}|^2}{t^2},\\
\label{T4ya}
|T|^2& =& I_0+\lambda_{p_{1}} \lambda_{e_{2}} I_1, \\
I_0&=&G_E^{\,2} Y_1 + \tau_p\, G_M^{\,2} Y_2,\\
I_1&=& \tau_p (\,G_E G_M Y_3 + G_M^{\,2} Y_4),
\ea
where $t=q^2$, $\lambda_s=4((p_1q_1)^2-m_0^2\,m^2)$, $\lambda_{p_{1}}$ ($\lambda_{e_{2}}$) --
the degree of polarization of the initial proton (of the final electron).

Here the functions $Y_i$ ($i=1, \ldots 4$) defined as
\ba
\label{Y1}
Y_1&=&(p_+q_+)^2+q_+^2q_-^2,\\
\label{Y2}
Y_{2}&=&(p_+ q_+)^2-q_+^2(q_-^2+4 m_0^2),\\
\label{Y3}
-Y_{3}&=& 2\,\kappa\, m^2\,((p_+q_+)^2+q_+^2(q_-^2-4\,m_0^2\,))\,z^2,\\
\label{Y4}
Y_{4}&=& 2\, (m^2 p_+ q_+ -\kappa q_+^2)(\kappa p_+ q_+ - m_0^2\, q_+^2) z^2,~~~\\
&& z= ( \kappa ^{2} - m^2 m_0^2 )^{-1/2}\,, \;\kappa=q_{1}p_{2} .\nn
\ea
Expression (\ref{T4ya}) for $|T|^2$ can be written as
\ba
\label{Tp1e2}
&&|T|^2=I_0+\lambda_{p_{1}} \lambda_{e_{2}} I_1=I_0\,(1+\lambda_{e_{2}}\lambda_{e_{2}}^{f}).
\ea
Then the value of $\lambda_{e_{2}}^{f}$ in (\ref{Tp1e2}) is the longitudinal polarization degree
transferred from the initial proton to the final electron in the $e\vec p\to \vec e p$ process
\ba
\label{lambdaef}
&& \lambda_{e_{2}}^{f}=\lambda_{p_{1}}\frac{I_1}{I_0}=\lambda_{p_{1}}
\frac{\tau_p (G_E G_M Y_3 + G_M^{\,2} Y_4)}{G_E^{\,2} Y_1 + \tau_p G_M^{\,2} Y_2}.
\ea
Dividing the numerator and denominator in the last expression by $Y_1 G_M^{\,2}$ and introducing
the experimentally measured ratio $R\equiv\mu_p G_E/G_M$, we get
\ba
\label{lambdae}
\lambda_{e_{2}}^{f}=\lambda_{p_{1}}\, \frac{\mu_p \tau_p\, ( (Y_3/Y_1) R+ \mu_p (Y_4/Y_1))}
{ R^{\,2} + \mu_p^2 \,\tau_p\, (Y_2/Y_1) \,}.
\ea
Inverting relation (\ref{lambdae}), we obtain a quadratic equation with respect to $R$:
\ba
\alpha_0 R^2 -\alpha_1 R + \alpha_0 \alpha_3-\alpha_2=0
\label{Rp}
\ea
with the coefficients:
\ba
\label{alpha1}
&& \alpha_0=\lambda_{e_{2}}^f / \lambda_{p_{1}}, \; \alpha_1=\tau_p\, \mu_p\, Y_3/Y_1,\\
&& \alpha_2=\tau_p\, \mu_p^2\, Y_4/Y_1, \; \alpha_3=\tau_p\, \mu_p^2\, Y_2/Y_1.\nn
\ea
Solutions to equation (\ref{Rp}) have the form:
\ba
\label{Rsqrt}
R=\frac{\alpha_1 \pm \sqrt{\alpha_1^2-4\alpha_0(\alpha_0\alpha_3-\alpha_2)}}{2\alpha_0}.
\ea
They allow us to extract the ratio $R$ from the results of an experiment to measure the polarization
transferred to the electron $\lambda_{e_{2}}^f$ in the $e\vec p\to \vec e p$ process in the case
when the scattered electron moves in the direction of the spin quantization axis
of the initial resting proton.

In the ultrarelativistic limit, when the electron mass can be neglected,
expressions (\ref{Y1})--(\ref{Y4}) for $Y_i$ ($i=1, \ldots 4$) in LF are given by
\ba
\label{Y1a}
&&Y_1 =8m^2 (2 E_1E_2 - m E_- ), \\
\label{Y2a}
&&Y_2=8 m^2 (E_1^{\,2} + E_2^{\,2} + m E_-),\\
\label{Y3a}
&&Y_3=-(2m/E_2)\, Y_1 ,\\
\label{Y4a}
&&Y_4=8 m^2 E_+ E_- (m-E_2)/E_2,
\ea
where $E_{\pm}=E_1 \pm E_2$.

\begin{figure}[h!]
\centering
\hspace{-5mm}
\includegraphics[scale=0.45]{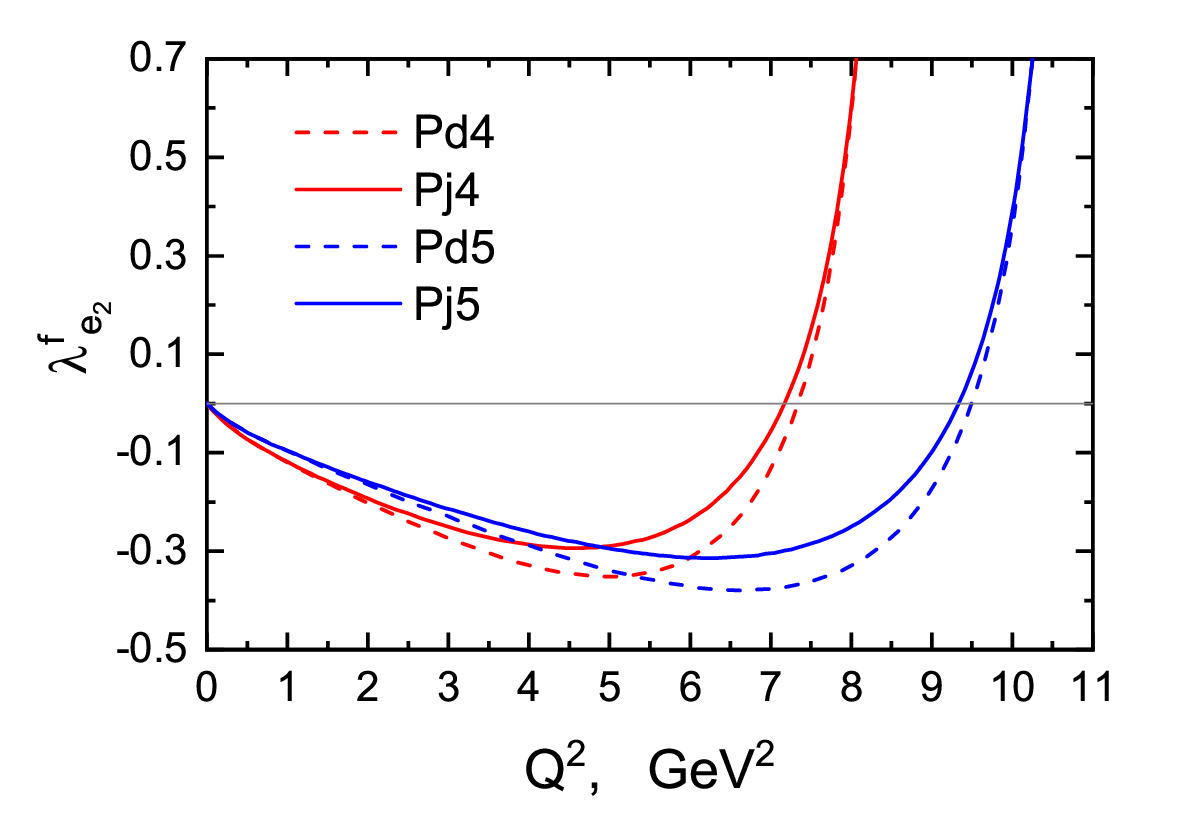} % 3 Mb
\vspace{-8mm}
\caption{
$Q^2$-dependence of the longitudinal polarization degree of the scattered electron
$\lambda_{e_{2}}^{f}$ (\ref{lambdae}) at electron beam energies in the experiment
\cite{Liyanage2020}. The lines $Pd4$, $Pd5$ (dashed) and $Pj4$, $Pj5$ (solid) correspond
to the ratio $R=R_d$ in the case of dipole dependence %($R_d=1$)
and parametrization $R=R_j$ (\ref{Rdj}) from the paper \cite{Qattan2015}.
The lines $Pd4$, $Pj4$ ($Pd5$, $Pj5$) correspond to the energies $E_1=4.725$ ($5.895$) $\GEV$.
}
\label{exp12}
\end{figure}

\begin{figure}[h!]
\centering
\includegraphics[width=0.45\textwidth]{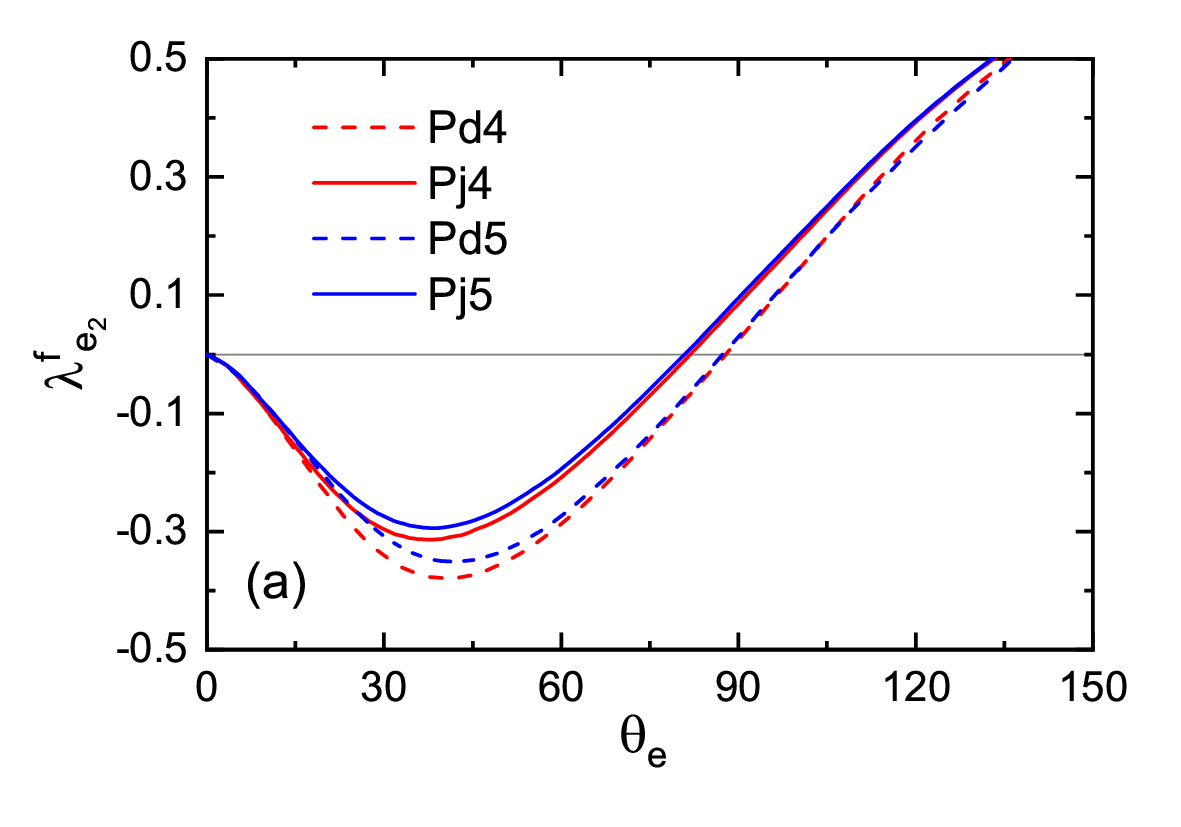}\\
\vspace{-5mm}
\includegraphics[width=0.45\textwidth]{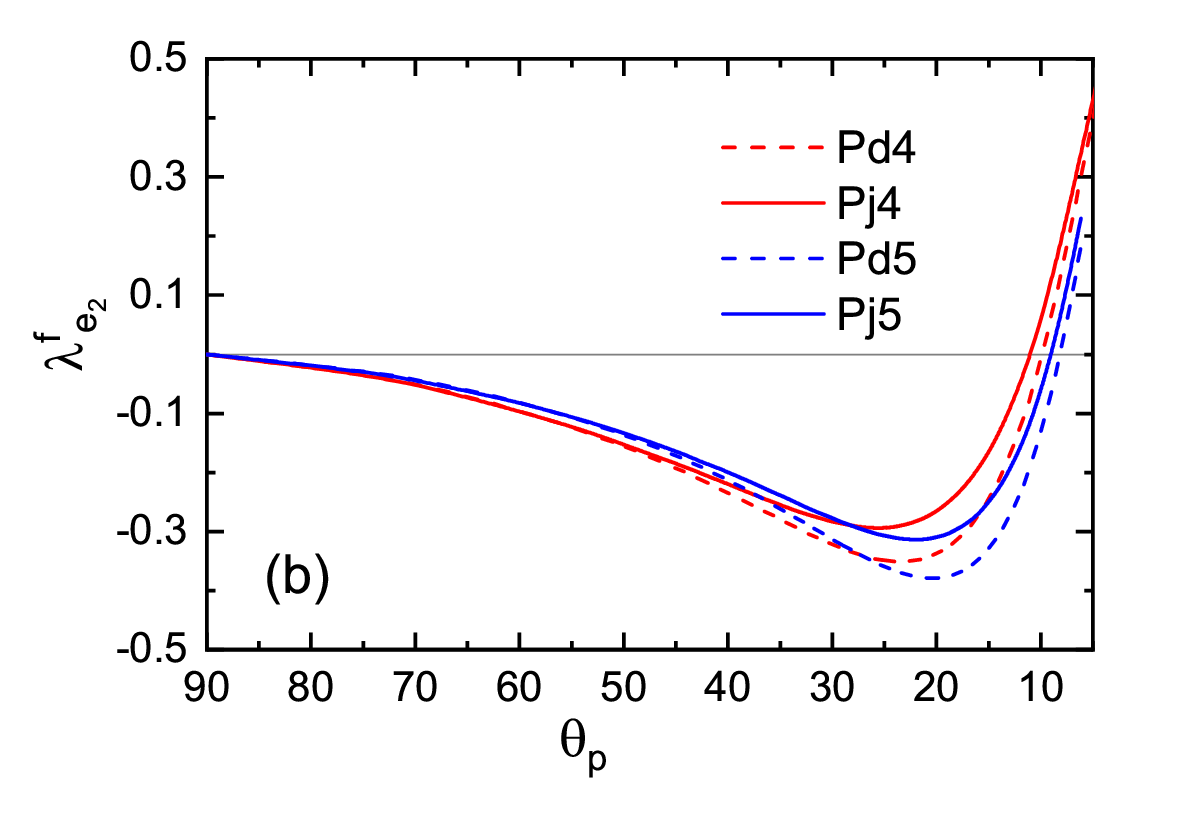} %\\
\vspace{-6mm}
\caption{
Angular dependence of the longitudinal polarization degree of the scattered electron
$\lambda_{e_{2}}^{f}$ (\ref{lambdae}) %in the $e\vec p\to\vec e p$ process
at electron beam energies in the experiment \cite{Liyanage2020}.
Panels (a) and (b) correspond to $\theta_e$-and $\theta_p$-dependencies expressed in degrees.
The marking of lines $Pd4$, $Pd5$, $Pj4$, $Pj5$ is the same as in Figure~\ref{exp12}.
}
\label{exp13}
\end{figure}

The formulas (\ref{Y1a})--(\ref{Y4a}) were used to numerically calculate the $Q^2$-dependence
of the longitudinal polarization degree of the scattered electron $\lambda_{e_{2}}^f$ (\ref{lambdae})
as well as the dependencies on the scattering angles of the electron and proton at electron
beam energies ($E_1=4.725$ and $5.895~\GEV$) and the polarization degree of the proton target
($P_t=\lambda_{p_{1}}=0.70$) in the SANE collaboration experiment \cite{Liyanage2020}
as while conserving the scaling of the SFF in the case of a dipole dependence ($R=R_d=1$),
and in case of its violation. In the latter case, the parametrization $R=R_j$ from
the paper \cite{Qattan2015} was used
\ba
R_j = (1+0.1430\,Q^2-0.0086\,Q^4+0.0072\,Q^6)^{-1},~~~
\label{Rdj}
\ea
and also the parametrization of Kelly from \cite{Kelly2004}, formulas for which ($R=R_k$) we omit.
The calculation results are presented by graphs in Figures~\ref{exp12}, \ref{exp13}.
Note that in these figures there are no lines corresponding to the parametrization of Kelly \cite{Kelly2004}
since calculations using $R_j$ and $R_k$ give almost identical results.

{\it Results of numerical calculations}.---$Q^2$-dependence of the longitudinal polarization
degree of the scattered electron $\lambda_{e_{2}}^{f}$
(\ref{lambdae}) at the electron beam energies in the experiment \cite{Liyanage2020} is presented
by graphs in Figure~\ref{exp12}, on which the lines $Pd4$, $Pd5$ (dashed) and $Pj4$, $Pj5$ (solid)
are constructed for $R=R_d$ and $R=R_j$ (\ref{Rdj}). At the same time, the red lines $Pd4$, $Pj4$
and the blue lines $Pd5$, $Pj5$ correspond to the energy of the electron beam $E_1=4.725$
and $5.895~\GEV$. For all lines in Figure~\ref{exp12} the degree of polarization of the proton target $P_t=0.70$.
\vspace{-2mm}

As can be seen from Figure~\ref{exp12}, the function $\lambda_{e_{2}}^{f}(Q^2)$ (\ref{lambdae})
takes negative values for most of the allowed values and has a minimum for some of them.
%we will indicate them as $Pd4(4.976)=-0.352$, $Pj4(4.586)=-0.294$,
%$Pd5(6.648)=-0.380$, $Pj5(6.254)=-0.314$.
On a smaller part of the allowed values adjacent
to $Q^2_{max}$ and amounting to approximately 9\% of $Q^2_{max}$, it takes on positive values.
At the boundary of the spectrum at $Q^2=Q^2_{max}$, the polarization transferred to the electron is
equal to the polarization of the proton target, $\lambda_{e_{2}}^{f}(Q^2_{max})=P_t=0.70$.

\begin{table*} [h!tpb]
\centering
\caption{
The degree of longitudinal polarization of the scattered electron $\lambda_{e_{2}}^{f}$ (\ref{lambdae})
at electron beam energies $E_1=4.725$ and $5.895~\GEV$ and two values $Q^2=2.06$ and $5.66~\GEV^2$
in the experiment \cite{Liyanage2020}. The values in the columns for $P_d$, $P_j$, $P_k$ correspond
to the polarization transferred to the electron $\lambda_{e_{2}}^{f}$ (\ref{lambdae})
with dipole dependence, the parametrization (\ref{Rdj}) of Qattan \cite{Qattan2015}
and Kelly \cite{Kelly2004}. The corresponding electron
and proton scattering angles (in degrees) are given in columns for $\theta_{e}$ and $\theta_{p}$.
}
\vspace{1mm}
\label{DeltaKelly}
\tabcolsep=2.90mm
\footnotesize
\begin{tabular}
{| c | c | c | c | c | c | c | c | c |}
\hline
$E_1$, $\GEV$
&  $Q^2$, $\GEV^2$
& $\theta_{e}\, (^{\circ})$
& $\theta_{p} \,(^{\circ})$
& $P_d$
& $P_j$
& $P_k$
& $\Delta_{dj}$, \%
& $\Delta_{jk}$, \% \\
\hline
5.895 &2.06 & 15.51 & 45.23 & --0.170 & --0.163 & --0.163 & 4.1 & 0.0 \\
\hline
5.895 &5.66 & 33.57 & 24.48 & --0.363 & --0.309 & --0.308 & 14.9  & 0.3 \\
\hline
4.725 &2.06 & 19.97 & 43.27 & --0.207 & --0.197 & --0.197 & 4.8 & 0,0 \\
\hline
4.725 &5.66 & 49.50 & 19.77 & --0.336 & --0.263 & --0.262 & 21.7 & 0.6\\
\hline
\end{tabular}
\end{table*}

Figure~\ref{exp13} shows the angular dependence
of the transferred to the electron polarization $\lambda_{e_{2}}^{f}$ (\ref{lambdae})
in the $e\vec p\to \vec e p$ process at electron beam energies $4.725$ and $5.895~\GEV$
in the experiment \cite{Liyanage2020}. The degree of polarization of the proton target was taken
the same for all lines: $P_t=0.70$. Panels (a) and (b) correspond to the dependence
on the scattering angles of the electron ($\theta_e$) and proton ($\theta_p$),
expressed in degrees.

\begin{figure}[h!tpb]
\centering
\includegraphics[width=0.45\textwidth]{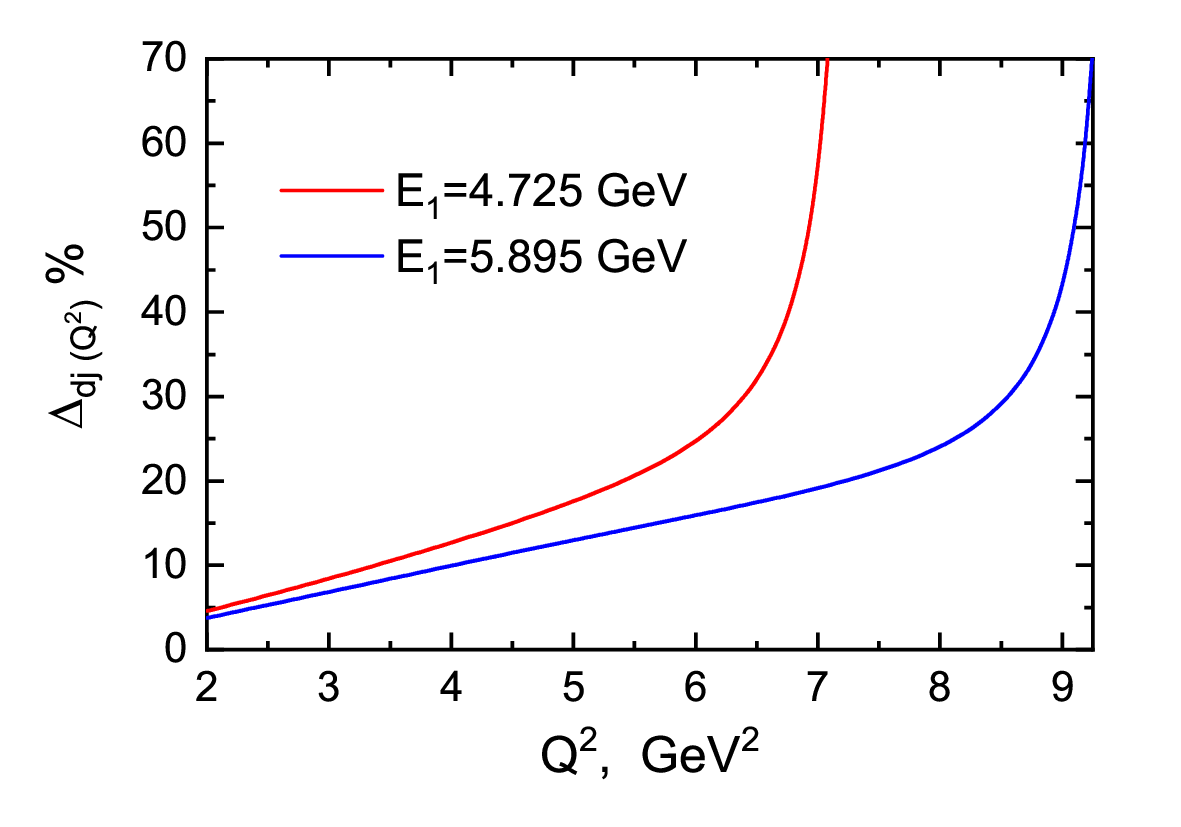}
\vspace{-5mm}
\caption{
$Q^2$-dependence of the relative difference $\Delta_{dj}$ (\ref{Deltadj}) at electron
beam energies $E_1=4.725~\GEV$ (red line) and $E_1=5.895~\GEV$ (blue line).
For all lines, the degree of polarization of the proton target was taken to be the same $P_t=0.70$.
}
\label{exp14}
\end{figure}

The parametrizations of Qattan \cite{Qattan2015} and Kelly \cite{Kelly2004} allow us
to calculate the relative difference $\Delta_{dj}$ between the polarization effects
in the $e\vec p\to\vec e p$ process in the case of conservation and violation of the SFF scaling,
as well as in the effects between these parametrizations $\Delta_{jk}$:
\ba
\label{Deltadj}
\Delta_{dj}=\Big|\frac{\rm{Pd}-\rm{Pj}}{\rm{Pd}}\Big|, \;
\Delta_{jk}=\Big|\frac{\rm{Pj} - \rm{Pk}}{\rm{Pj}}\Big|,
\ea
where $P_d$, $P_j$ and $P_k$ are the polarizations calculated by formula~(\ref{lambdae}) for
$\lambda_{e_{2}}^{f}$ when using the corresponding parametrizations $R_d$, $R_j$ and $R_k$.
The results of calculations of $\Delta_{dj}$ at electron beam energies of $4.725$ and $5.895~\GEV$ are
shown in Figure~\ref{exp14}.

It follows from the graphs in Figure~\ref{exp14} that the relative difference between the polarization
transferred from the initial proton to the final electron in the $e\vec p\to \vec e p$ process
in the case of conserving and violation of the scaling of the SFF can reach 70\%, which can be used
to set up a polarization experiment by measuring the ratio $R$.

Numerical values of the polarization transferred to the final electron in the $e\vec p\to\vec e p$
process for the three considered parametrizations of the ratio $R$ at $E_1$ and $Q^2$ used
in the experiment \cite{Liyanage2020}, are presented in Table~\ref{DeltaKelly}. In it, the columns
of values $P_d$, $P_j$ and $P_k$ correspond to the dipole dependence $R_d$, parametrizations
$R_j$ (\ref{Rdj}) and $R_k$ \cite{Kelly2004}; columns $\Delta_{dj}$, $\Delta_{jk}$ correspond
to the relative difference (\ref{Deltadj}) (expressed in percent) at electron beam energies
of $4.725$ and $5.895~\GEV$ and two values of $Q^2$ equal to $2.06$ and $5.66~\GEV^2$. It follows
from Table~\ref{DeltaKelly} that the relative difference between $Pj5$ and $Pd5$ at $Q^2=2.06~\GEV^2$
is 4.1\% and between $Pj4$ and $Pd4$ it is 4.8\%. At $Q^2 = 5.66~\GEV^2$, the difference increases
and becomes equal to 14.9 and 21.7\%, respectively. Note that the relative difference $\Delta_{jk}$
between $P_j$ and $P_k$ for all $E_1$ and $Q^2$ in Table~\ref{DeltaKelly} is less than 1\%.

\vspace{-2mm}

{\it Conclusion}.---In this paper, we have considered a possible method for measuring the ratio
$R \equiv \mu_p G_E/G_M$ based on the transfer of polarization from the initial proton to the final
electron in the $e\vec p\to \vec e p$ process, in the case when their spins are parallel,
i.e. when an electron is scattered in the direction of the spin quantization axis of the resting
proton target.
For this purpose, in the kinematics of the SANE collaboration experiment \cite{Liyanage2020},
using the parametrizations of Qattan \cite{Qattan2015} and Kelly \cite{Kelly2004}, a numerical
analysis was carried out of the dependence of the degree of polarization of the scattered
electron on the square of the momentum transferred to the proton, as well as from the scattering
angles of the electron and proton.
As it turned out, the parametrizations of Qattan \cite{Qattan2015} and Kelly \cite{Kelly2004} give
almost identical results in calculations.
It is established that the difference in the degree of longitudinal polarization of the final
electron in the case of conservation and violation of the SFF scaling can reach 70 \%,
which can be used to conduct a new type of polarization experiment to measure the ratio $R$.

At present, an experiment to measure the longitudinal polarization degree transferred
to an unpolarized electron in the $e\vec p\to \vec e p$ process when it is scattered in
the direction of the spin quantization axis of a resting proton seems quite real
since a proton target with a high degree of polarization $P_t = 70 \pm 5$ \% was
created in principle and has already been used in the experiment \cite{Liyanage2020}.
For this reason, it would be most appropriate to conduct the proposed experiment
at the setup used in \cite{Liyanage2020} at the same $P_t=0.70$, electron beam energies
$E_1=4.725$ and $5.895~\GEV$. The difference between conducting the proposed experiment and
the one in \cite{Liyanage2020} consists in the fact that an incident electron beam must
be unpolarized, and the detected scattered electron must move strictly along the direction
of the spin quantization axis of the proton target. In the proposed experiment, it is necessary
to measure only the longitudinal polarization degree of the scattered electron, which
is an advantage compared to the method \cite{Rekalo74} used in JLab-experiments.

{\it Acknowledgements}.---This work was carried out within the framework of scientific cooperation
Belarus-JINR and State Program of Scientific Research ``Convergence-2025'' of the Republic of Belarus
under Projects No. 20221590 and No. 20210852.

\end{document}